\documentclass[journal, letterpaper]{IEEEtran}
\usepackage{amsmath}
\usepackage{comment}
\usepackage[table]{xcolor}
\usepackage{url} 
\usepackage{dsfont}
\usepackage{cite,cleveref}
\usepackage{tabularx}
\usepackage{subcaption}
\usepackage{graphicx}
\usepackage{soul}
\usepackage{balance}
\usepackage{xspace}
\usepackage{color}
\usepackage{booktabs}
\usepackage{todonotes}
\begin{document}

\title{Application-Integrated Slicing towards 6G:\\The Musical Metaverse Use Case\thanks{This work is supported by MUSMET project funded by the EIC Pathfinder Open scheme of the European Commission (grant agreement n. 101184379), and has received funding from the
Swiss State Secretariat for Education, Research and Innovation (SERI). Corresponding author: ali.alhousseini@supsi.ch}}

\author{  
\IEEEauthorblockN{
Ali Al Housseini\IEEEauthorrefmark{1}, 
Jaime Llorca\IEEEauthorrefmark{2}\IEEEauthorrefmark{4}, 
Omran Ayoub\IEEEauthorrefmark{1}, 
Cristina Rottondi\IEEEauthorrefmark{3}, 
Luca Turchet\IEEEauthorrefmark{2}, 
Francesco Malandrino\IEEEauthorrefmark{5}\IEEEauthorrefmark{6}
}\\
\IEEEauthorblockA{\IEEEauthorrefmark{1}
University of Applied Sciences and Arts of Southern Switzerland, Lugano, Switzerland}
\\
\IEEEauthorblockA{\IEEEauthorrefmark{2}
University of Trento, Trento, Italy;}
\IEEEauthorblockA{\IEEEauthorrefmark{3}
Politecnico di Torino, Turin, Italy}
\\
\IEEEauthorblockA{\IEEEauthorrefmark{4}
Centre Tecnologic de Telecomunicacions de Catalunya (CTTC/CERCA), Castelldefels, Spain.}
\\
\IEEEauthorrefmark{5} CNR-IEIIT, Italy;\IEEEauthorrefmark{6} CNIT, Italy
}

\maketitle

\begin{abstract}
Emerging immersive applications are expected to support heterogeneous groups of users with fundamentally different communication and computation requirements. Using the Musical Metaverse (MM) as a representative example, we show how such applications expose limitations of current 5G network slicing and orchestration frameworks, which remain largely service-centric and operate through decoupled application and network management mechanisms. While existing Quality of Service (QoS) and slicing techniques provide traffic differentiation, they cannot jointly account for application semantics, shared resources, and differentiated end-to-end requirements across multiple user classes within the same service instance. To address these limitations, we introduce \emph{application-integrated slicing}, a framework that unifies application and network orchestration within a common end-to-end service model and enables differentiated KPI targets for multiple user classes within a single logical slice. Using a MM reference scenario deployed over cloud--edge network, we show that application-integrated slicing reduces both resource provisioning cost and QoS violation rates compared with conventional decoupled approaches. 
\end{abstract}

\section{Introduction}
Fifth-generation (5G) mobile networks have introduced the paradigm of network slicing, enabling the creation of virtualized logical networks tailored to specific service classes~\cite{foukas2017slicing,3gpp23501}. 
Within a slice, further differentiation is possible through mechanisms such as Quality of Service (QoS) flows, traffic marking, and sub-slicing, which allow scheduling priority and radio resource partitioning to vary across user groups. However, these mechanisms remain scoped to the network transport layer and the delay bounds they enforce apply between the user device and the network edge, with no control over application function placement or the compute delays incurred beyond it. Consequently, even a slice that satisfies its transport-level delay bound cannot guarantee the end-to-end latency experienced by the user, since a substantial share of that latency arises from function placement and inter-function routing that lie outside the slice's control.

While mechanisms exist to differentiate user experiences, both at the network level (e.g., QoS flows, sub-slicing, traffic marking), and at the application level (e.g., microservice placement, service mesh traffic policies), these operate in a largely {\em decoupled} manner, across distinct domains and time scales, with no shared end-to-end service model. 

This decoupled approach has proven effective for conventional 5G service classes where service requirements are relatively homogeneous. Under these conditions, service-level KPI targets can be effectively met through decoupled orchestration and resource management mechanisms, without requiring tight cross-domain coordination~\cite{3gpp23501}. 
However, emerging 6G applications are expected to support heterogeneous groups of users and services with fundamentally different requirements. 
In such settings, satisfying aggregate service-level KPIs is no longer sufficient, as different user classes may experience markedly different levels of service quality.
The absence of a shared end-to-end service model limits the ability of independent orchestration layers to enforce per-class guarantees. Specifically, the application orchestrator cannot control how network resources and policies affect each user class traffic, while the network orchestrator cannot account for how shared application-functions and compute resources are distributed across users. 

The musical metaverse (MM), introduced in \Cref{sec:reqs}, exemplifies this emerging class of services. MM integrates immersive metaverse environments with real-time networked music performance, enabling geographically distributed performers (i.e., musicians) and audiences to interact within a shared virtual space under stringent synchronization and latency requirements. As such, it combines interactive, heterogeneous, and multicast-oriented communication patterns that demand tightly coordinated decisions across both application and network planes~\cite{dionisio2013metaverse}.
Such applications involve users with distinct roles, priorities, and performance expectations that are simultaneously coupled through shared application and network resources. Meeting their requirements therefore necessitates tightly coordinated orchestration decisions across both the application and network domains.

\begin{figure*}[t]
\centering
\includegraphics[width=0.9\linewidth]{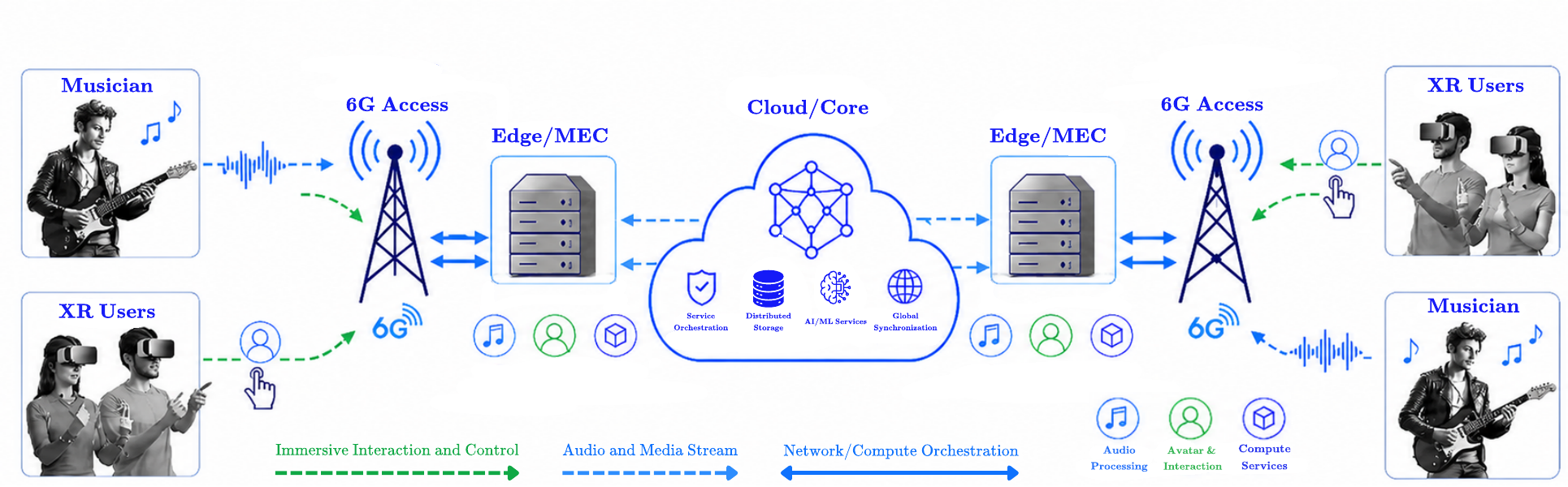}
\caption{High-level architecture of a MM session illustrating how distributed musicians and XR users exchange audio/media streams and control signals over a 6G edge–cloud network.}
\label{fig:musmet_example}
\end{figure*}

\Cref{fig:musmet_example} illustrates a representative MM session involving multiple user classes with distinct roles and service requirements. Professional musicians generate and exchange latency-sensitive audio streams to maintain musical synchrony, while audience members consume immersive audiovisual content with less stringent timing constraints. These interactions rely on shared application functions and network resources, creating complex dependencies between users with heterogeneous performance objectives. 
Consequently, resource allocation decisions at both the application and network levels directly influence the quality of experience perceived by each user class~\cite{dionisio2013metaverse,chafe2004latency}.
Despite sharing the same service environment, users in such applications exhibit fundamentally different performance requirements. Performers require ultra-low latency and highly deterministic communication to preserve rhythmic synchrony, whereas spectators prioritize immersive rendering quality and can tolerate higher end-to-end delays. These user roles interact within the same session, consume related data streams, and must remain temporally aligned, further complicated by heterogeneous devices and multimodal data flows.

Current 5G mechanisms address such heterogeneity primarily by assigning users to different QoS flows and prioritizing traffic accordingly. However, these mechanisms lack the granularity and scalability required by emerging applications such as the MM. In these environments, tightly coupled, multimodal interactions require network-level decisions that directly reflect application-level semantics, rather than relying on indirect mappings or over-provisioning.
These challenges expose a fundamental architectural limitation: existing network slicing and application orchestration mechanisms operate as \emph{independent} layers, each optimizing its own domain without a shared end-to-end service model. 

In this paper, we introduce {\em application-integrated slicing}: a framework that addresses this limitation by jointly optimizing application microservices and network functions within a unified end-to-end service model that captures both application semantics (with granular per-class KPI targets) and matching network functionality. Rather than treating application and network orchestration as separate problems, application-integrated slicing extends the network orchestrator's scope to include application functions and per-class/role flows as co-optimizable variables, enabling role-aware, end-to-end resource management within a single logical slice. Using the MM as a representative and demanding case study, we demonstrate that application-integrated slicing reduces both resource provisioning cost and per-class KPI violation rates compared to decoupled orchestration, with gains that become substantial as session heterogeneity and load increase.

\begin{figure*}[t]
    \centering
    \includegraphics[width=0.7\linewidth]{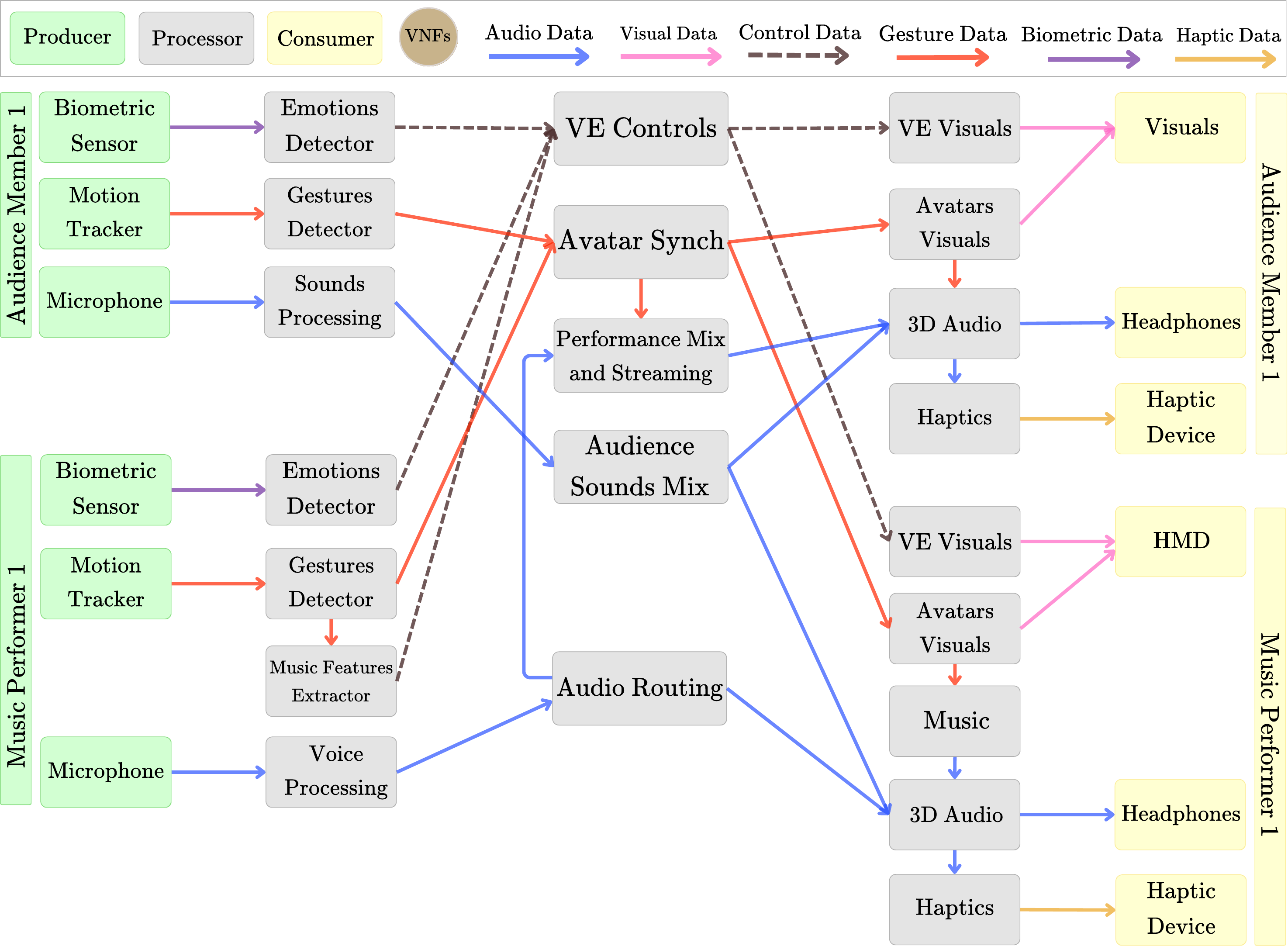}
    \caption{Functional service graph representation of the MM reference scenario.}
    \label{fig:service_graph}
\end{figure*}

\section{Technical requirements for the musical metaverse} 
\label{sec:reqs}

The MM represents a new class of immersive systems where musicians and audiences interact in shared virtual environments to create, perform, and experience music in real time. Unlike current eXtended Reality (XR) applications, it aims to enable synchronous, multisensory, and interactive experiences that capture not only audio-visual content but also gestures, context, and even users' emotional states~\cite{dionisio2013metaverse}. Such systems have the potential to transform how music is composed, performed, and consumed, while also serving as a benchmark for real-time collaborative applications in the broader metaverse~\cite{turchet2023musical}. Achieving this vision, however, requires overcoming stringent technical challenges in latency, synchronization, scalability, and multimodal data processing that go well beyond those of existing XR and gaming platforms.

{\bf Beyond immersive gaming}.
Immersive musical interaction imposes constraints that are fundamentally different from those of mainstream metaverse applications such as immersive gaming. While modern XR platforms target perceptual responsiveness and overall quality of experience, the MM requires \emph{body-level synchronization} across distributed participants, where timing errors directly impact human motor coordination and collective performance.
This distinction becomes evident when considering the coexistence of performers and audience members within the same session: in immersive gaming, users typically share similar interaction models and operate under comparable latency requirements, often tolerating end-to-end delays on the order of 50--100 ms through prediction and compensation techniques \cite{rottondi}. Musical interaction, by contrast, does not admit such compensation: prediction and buffering rely on a degree of perceptual tolerance that is absent when timing itself carries the musical content rather than merely conveying delayed events.

{\bf Role asymmetry: performers vs audience}.
A performer engaged in real-time collaboration requires end-to-end latencies below approximately 20--30 ms to preserve rhythmic synchrony; beyond this threshold, temporal misalignment degrades ensemble coordination. Unlike typical XR applications, these interactions rely on precise real-time feedback loops that cannot be effectively predicted or buffered. Audio streams must therefore follow deterministic paths with minimal jitter, while gesture tracking and spatial positioning must remain aligned with the audio stream within a few milliseconds.

By contrast, audience members prioritize immersion quality over ultra-low latency. Spatial audio, high-resolution visuals, and environmental effects dominate their experience, and end-to-end latencies in the range of 80--150 ms remain acceptable as long as audiovisual coherence is preserved. This asymmetry creates a core architectural challenge: simultaneously supporting ultra-low-latency communication for performers and bandwidth-intensive, latency-tolerant delivery for spectators within the same session.

This challenge is further compounded by device diversity. Performers often rely on specialized low-latency equipment such as digital musical instruments or edge-connected systems, whereas spectators access the experience through head-mounted displays or cloud-rendered XR clients. These platforms differ significantly in computing capabilities, network connectivity (performance, stability, availability), and rendering pipelines; all such diversity requires adaptive resource management across the system.

{\bf Multimodal coupling and synchronization constraints}.
In addition to role differentiation, immersive musical sessions generate multiple concurrent data flows with distinct and interdependent timing constraints. A single session may include any of all the following flows:
\begin{itemize}
\item real-time musical audio ($<$30 ms latency requirement);
\item avatar motion and gesture tracking (frame-synchronous, 60--90 Hz);
\item spatial audio rendering (per-frame updates);
\item visual scene updates (typically 90 Hz in VR systems);
\item optional neurophysiological or engagement signals (10--100 Hz sampling, with higher latency tolerance).
\end{itemize}

While similar flows exist in other XR applications, their coupling is significantly tighter in musical interaction. Perceptual coherence imposes strict cross-modal synchronization constraints: for performers, audio--gesture misalignment beyond 10--15 ms disrupts coordination, whereas for spectators audiovisual offsets above 40--60 ms become noticeable. These constraints must therefore be satisfied jointly across all streams rather than treated independently.
This combination of role asymmetry, device diversity, and tightly coupled multimodal constraints defines the core systems challenge of the MM. Supporting such requirements demands joint, dynamic, role-aware adaptation of network paths, compute placement and rendering pipelines, while preserving temporal coherence across the entire session.

\begin{figure*}
\centering
\includegraphics[width=0.8\textwidth]{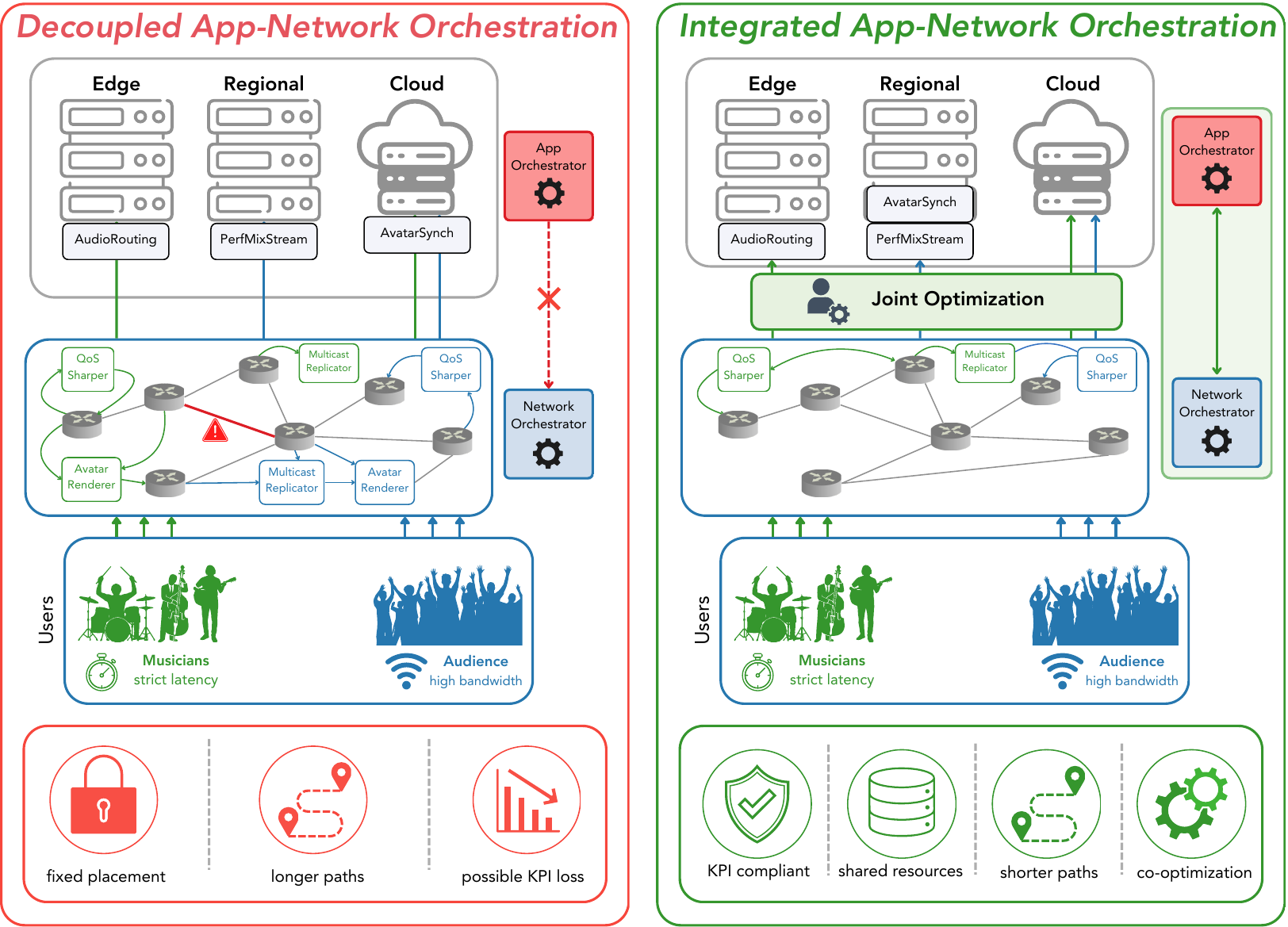}
\caption{Comparison of decoupled and application-integrated orchestration for an MM session. Left: application and network orchestrators operate independently. Right: a joint optimization layer jointly manages placement, routing, and resources across both domains to improve efficiency and KPI compliance.}
\label{fig:decoupledVSjoint}
\end{figure*}

\section{Reference Scenario}
To ground our discussion, we present a concrete MM application as reference scenario. 
This scenario represents a challenging application, combining two user classes with intrinsically different requirements within the same session: a relatively small set of musicians and a much larger set of audience members. Moreover, the scenario integrates real-time interaction among geographically distributed musicians performing with virtual musical instruments and
a distributed VR audience attending the concert and interacting in the virtual environment.
As discussed in \Cref{sec:reqs}, musicians require stringent latency and jitter control for synchronous co-performance, while audience members prioritize stable, high-quality multimodal delivery.

{\bf Two-plane service architecture}.
The corresponding service graph, illustrated in \Cref{fig:service_graph}, is organized around two tightly coupled planes. The \emph{performer interaction plane} supports real-time musician-to-musician coupling through sound-producing gestures and low-latency voice exchange. The \emph{audience program plane} generates and distributes a coherent show stream, including rendered music, voice, avatars, and environmental context. 
This separation is key: musicians rely on ultra-responsive interaction driven by synchronized events, whereas audience members consume a buffered and quality-stabilized program stream.

{\bf Functional components}. From a functional perspective, our scenario includes: $(i)$ per-user producers and consumers (e.g., biometrics, gestures, contextual sounds, voice encoding) and $(ii)$ shared processors for synchronization and aggregation (e.g., Avatar Synch, Audio Routing, Audience Sounds Mix, Performance Mix and Streaming, VE Controls).

The scenario is naturally designed as directed-acyclic graph of \emph{producers} (sources), \emph{processors} (in-network functions), and \emph{consumers} (destinations), connected by edges carrying distinct modalities, each with its own rate and latency profile. Producers capture per-participant input: both roles generate biometric and motion data, while audience members add contextual \lq \lq crowd'' sounds and musicians add voice and high-frequency \emph{sound-producing gestures}. Consumers render the result through head-mounted displays, spatial audio, and haptics. Processors perform the in-network computation, from per-user encoding and rendering to core shared functions such as \emph{Avatar Synch}, \emph{Audio Routing}, \emph{Audience Sounds Mix}, and \emph{Performance Mix and Streaming}.

Two design choices make this graph far more tightly coupled than a conventional XR session. First, musicians transmit \emph{gestures rather than audio}, with sound synthesized locally at each peer cutting bandwidth but sharply raising sensitivity to latency and jitter. Second, shared functions serve both classes under conflicting budgets: \emph{Avatar Synch}, for instance, feeds a single gesture stream into both a tight performer loop and the audience program path, while delivery itself is asymmetric, the buffered program mix reaches the audience only, as musicians stay on a separate low-latency loop. It is this combination of synchronized multimodal flows, shared role-serving functions, and asymmetric delivery that makes the MM a demanding benchmark for joint application--network orchestration.


\section{Current approaches}

This section reviews current 5G slicing, orchestration, and scheduling approaches, and their limitations for emerging services such as the MM.

\subsection{Slicing, orchestration, and scheduling in 5G}
\label{sec:current}

Resource management in 5G networks is structured across three main layers operating at different time scales: network slicing, orchestration, and scheduling. These layers collectively aim to ensure that service-level Key Performance Indicators (KPIs) are met under dynamic network conditions.

At the longest time scale, \emph{network slicing} defines logical network instances tailored to specific services. Each slice is associated with a well-defined set of KPIs -- such as latency, reliability, and throughput -- which are assumed to apply uniformly to all users of that service. This establishes a fundamental design principle: \emph{one service, one slice, one set of KPIs}. While extensions such as sub-slicing or network slice subnet instances allow further partitioning of resources, they still inherit this service-centric view, where KPIs are defined per service rather than per user or role.

At intermediate time scales, \emph{network orchestration} is responsible for mapping slices onto physical and virtual resources, including compute, storage, and networking components. Orchestration dynamically adapts to variations in traffic demand, user distribution, and resource availability, with the goal of maintaining slice-level KPIs. This often involves reallocating resources across virtual network functions (VNFs), or moving the VNFs themselves to another node within the network infrastructure.

At the shortest time scale, \emph{scheduling} operates at the radio and transport levels, assigning resources to individual users and flows. Scheduling mechanisms can differentiate between users through priority levels, QoS classes, and traffic descriptors. 
However, 
due to the lack of native awareness of application-level roles (such as performer or audience), enforcing role-specific end-to-end guarantees that span both network and application domains remains beyond the reach of scheduling mechanisms alone.


\subsection{Limitations of current approaches}

Current approaches and algorithms have proven effective for services characterized by relatively homogeneous requirements, such as the ones prevalent in present-day 5G networks. However, they show fundamental limitations when applied to emerging applications such as the MM where users within the same service may have widely different performance needs (as described in \Cref{sec:reqs}).

At its core, network slicing is designed to guarantee that service-level KPIs are met, i.e., that any request associated with a given service satisfies predefined latency, bandwidth, or reliability targets. Orchestration ensures that sufficient resources are allocated to each slice to uphold these guarantees under changing conditions. Scheduling, in turn, provides limited user-level differentiation but remains constrained to transport-level abstractions and coarse QoS classes.

Combined together, such approaches make it inherently difficult to accommodate heterogeneous per-class-role requirements end to end. 


More fundamentally, current approaches lack the ability to {\em jointly} consider {\em application-level} semantics and {\em network-level} resource allocation. As discussed in \Cref{sec:reqs}, emerging immersive applications require coordinated, role-aware decisions that span both domains, rather than isolated optimizations. 

These limitations motivate the need for a new decision-making approach to the orchestration problem. In the following section, we argue that 6G networks should move beyond service-centric slicing and support multiple classes of users with differentiated KPIs within the same logical slice. To this end, we introduce the concept of \emph{application-integrated slicing}, which integrates user-level differentiation with joint application–network resource management.

\section{The application-integrated slicing approach}
\label{sec:idea}

The proposed application-integrated slicing approach is illustrated in \Cref{fig:fig4} and is built around three main pillars:
\begin{itemize}
	\item Joint \emph{representation} of application and network resources, integrating application microservices and network functions for joint end-to-end orchestration.
	\item Enhanced \emph{algorithms} for end-to-end optimization and control decisions, enabling joint VNF–microservice placement, routing, scheduling, and resource allocation, that minimize overall resource consumption, while satisfying granular per role/class QoS constraints.  
	\item Multiscale \emph{implementation} capabilities for integrated network-application orchestrators that incorporate both long-term optimization as well as short-term control policies at different time and space scales.
\end{itemize}
We discuss each pillar below.

\subsection{Pillar I: Representation}

\begin{figure}[t]
\centering

\begin{subfigure}[t]{0.48\textwidth}
    \centering
    \includegraphics[width=\textwidth]{list_of_figures/fig3av3.pdf}
    \caption{Application-view service graph of the reference scenario, highlighting the coexistence of a performer interaction plane and an audience program plane.}
    \label{fig:scenario_sg4a}
\end{subfigure}
\hfill
\begin{subfigure}[t]{0.48\textwidth}
    \centering
    \includegraphics[width=\textwidth]{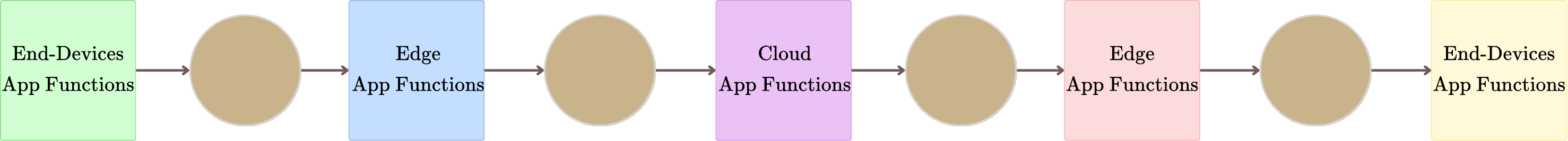}
    \caption{Network-view of the service graph model of the reference scenario.}
    \label{fig:scenario_sg4b}
\end{subfigure}
\caption{Service graphs associated with the distinct application and network views, representative of current decoupled network-application orchestration solutions, for the reference scenario.}
\label{fig:scenario_sg4}
\end{figure}

From the perspective of the network orchestrator, services are modeled through VNF forwarding graphs (VNFFGs)~\cite{houidi2020dynamic}, where vertices represent VNFs and edges their functional dependencies. Vertices and edges are annotated with computational and traffic requirements, and service-level KPIs, most notably end-to-end delay, guide orchestration decisions. 
However, VNFFGs do not capture granular application semantics, including per-class/role multimodal flows, multiple-input multiple-output application functions, and differentiated QoS requirements. This limited representation is illustrated in \Cref{fig:scenario_sg4b}.

On the other hand, from the perspective of the application orchestrator, applications are described via directed acyclic service graphs that capture application granularity \cite{mauro}, but ignore the VNFs required for ensuring end-to-end per-role requirements, as illustrated in \Cref{fig:scenario_sg4a}.

We extend this representation by explicitly modeling:
\begin{itemize}
	\item Application microservices and VNFs as differentiated vertices of the same service graph. 
	\item Per-role/class flows as differentiated edges within the same service graph.
\end{itemize}

\begin{figure*}[t]
    \centering
    \includegraphics[width=0.8\linewidth]{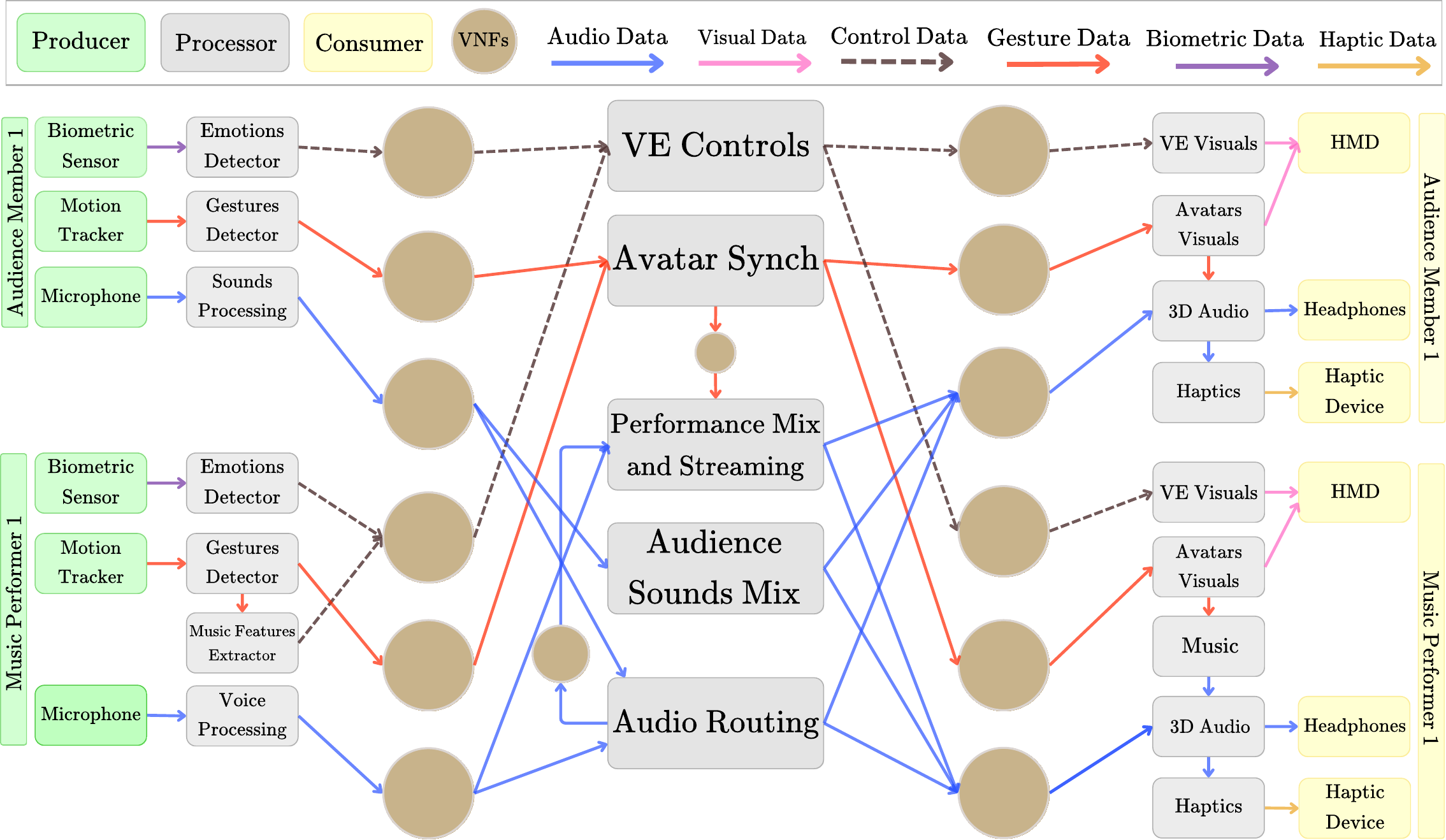}
    \caption{Service graph associated with the integrated application--network view, representative of the proposed application-integrated slicing solution, for the MM reference scenario. Application microservices and network functions (VNFs) are represented within a single service graph, with per-role flows modeled as distinct edges. The unified representation captures shared functions, asymmetric performer/audience delivery, and differentiated KPI requirements, enabling joint role-aware optimization within a single slice.}
    \label{fig:fig4}
\end{figure*}

This unified representation enables the orchestrator to reason jointly about application and network function placement, routing, and resource allocation under heterogeneous, per-class end-to-end constraints. The resulting graph structure, illustrated in \Cref{fig:fig4} forms the basis for the joint optimization algorithms described in Pillar II.

{\bf Separate application and network views}.
The two views in \Cref{fig:scenario_sg4a} (application-view) and \Cref{fig:scenario_sg4b} (network-view), respectively, represent the same scenario at different abstraction levels. The former shows the \emph{application service-graph}, highlighting functional blocks and application semantics: per-user producers, shared processing functions, and musician/audience consumers. 
The latter provides the \emph{network} view, with a much coarser representation of the application semantics and the addition of deployable VNFs. 
As described in the previous sections, this separate representation of application and network views, along with the resulting decoupled orchestration solutions constitute the main limitations of current slicing and orchestration approaches.

{\bf Integrated application-network view}.
The integrated application-network view, illustrated in \Cref{fig:fig4}, superimposes these two abstractions into a single service graph in which application microservices and deployable VNFs appear as vertices of the same structure. Every application-level dependency is realized as a path that may traverse one or more VNFs interleaved along the flow, so that the network realization of each function and inter-function flow becomes explicit and co-optimizable with microservice placement, rather than being deferred to a separate orchestration layer. Per-role flows are modeled as distinct edges, allowing differentiated KPI budgets to be attached to the specific paths that carry each class's traffic. Crucially, shared functions thereby become explicit objects of optimization: a processor consumed by both musicians and audience appears as a single vertex traversed by edges with different latency budgets, exposing a coupling that the separate views obscure, namely that a placement or routing decision taken to satisfy one class directly perturbs the budget of the other. This makes visible the central trade-off between \emph{replicating} a function to isolate stringent flows and \emph{aggregating} flows to preserve sharing, a trade-off invisible in either view alone, and the one to exploit which the joint algorithms of Pillar~II are designed.


\subsection{Pillar II: Algorithms}

The representation of Pillar I requires more sophisticated orchestration algorithms.
In application-integrated slicing, 
the same service graph may simultaneously impose multiple sets of end-to-end latency constraints, one per user class, over a graph that contains shared application and network functions consumed by multiple user classes.
This coupling significantly expands the solution space. 
For instance, increasing the latency of a shared application function may require compensating reductions elsewhere to avoid constraint violations, where such constraints may be different depending on the specific user class. 

These challenges motivate the use of advanced optimization techniques, including graph-theoretic and network-information-flow based formulations, such as those based on the cloud network flow (CNFlow) framework \cite{mauro} that allow optimizing the replication vs aggregation of different user flows and associated shared application functions. 
Despite increased complexity, the new integrated graph structure remains exploitable, enabling efficient CNFlow-based algorithms  even in the application-integrated slicing setting.

\subsection{Pillar III: Implementation}

In application-integrated slicing scenarios, decision complexity and input volume can increase latency, potentially limiting the system's ability to react to dynamic conditions. Addressing this requires balancing decision quality and responsiveness.

We adopt two complementary strategies. First, we accelerate decision-making by trading optimality for speed, e.g., terminating optimization early once near-optimal solutions are found and reducing input dimensionality through aggregation. Second, we improve robustness by incorporating safety margins in resource allocation, allowing the system to remain within KPI bounds while new decisions are computed.

Overall, this approach integrates optimization and control:
\begin{itemize}
    \item long-term, system-wide decisions are handled through optimization;
    \item short-term, fine-grained adjustments are managed via control policies.
\end{itemize}

\section{Numerical Assessment}
\label{sec:setup}


\subsection{Experimental Setup}

This section presents preliminary numerical results that support the promise of application-integrated slicing. To this end, we compare two orchestration approaches: \emph{Decoupled App--Network Slicing}, where application placement and network resource allocation are optimized separately, and \emph{6G Application-Integrated Slicing}, where application functions, VNFs, routing, and communication/computation resources are jointly optimized. The implementation follows a graph-based orchestration formulation with explicit end-to-end delay constraints and joint communication/computation resource accounting under the CNFlow modeling abstraction~\cite{mauro}.

To reflect a distributed edge--cloud deployment, the MM service graph is embedded onto a physical substrate network representing a 
NextG
transport network interconnecting compute-capable cloud and edge sites~\cite{polese2023oran,etsi2020mec003,taleb2017mec}. The underlying network is modeled as a substrate graph, where nodes correspond to network sites (e.g., central offices, aggregation points, edge data centers, or cloud facilities) equipped with communication and computing resources, while links represent transport-network connections characterized by finite bandwidth and propagation-delay constraints. 


Abilene and GEANT~\cite{SNDlib10} are adopted as representative network substrates because they exhibit different scales and connectivity structures while remaining widely used benchmark topologies. The original topologies are subsequently enriched with cloud/core, regional-edge, and access-edge tiers, together with compute resources and user attachment points, to emulate realistic edge--cloud deployments. Consequently, the evaluation compares orchestration strategies over two distinct infrastructure realizations rather than over a single topology-specific configuration.

Specifically, nodes are ranked according to their topological centrality, computed over the delay-weighted substrate graph. Highly central nodes are labeled as cloud/core sites, representing locations with substantial sites with larger compute capacity and traffic aggregation capabilities. Nodes with intermediate centrality are designated as regional-edge sites, while peripheral nodes are classified as access-edge sites, where end users and service endpoints are directly connected. 
This hierarchical assignment reflects advanced telecom architectures, with cloud/core sites providing centralized computing resources and edge sites enabling low-latency service delivery.

Each tier is assigned a distinct resource profile. Cloud/core nodes provide the largest compute capacity but may incur longer access paths; regional-edge nodes provide intermediate capacity; and access-edge nodes provide limited compute resources but are closer to end users. The same tier assignment, resource capacities, and user-attachment procedure are used for both orchestration modes, ensuring a fair comparison that isolates the impact of the orchestration strategy from that of the underlying infrastructure.

The settings are generated by varying the MM session size. Each experiment instantiates one MM service graph with a number of musicians $M \in \{2,4,6,8,10\}$ and a number of audience members $A \in \{10,20,30\}$. Musicians represent interactive service endpoints that generate performance-related traffic, while audience members represent receiving endpoints. Increasing $M$ increases the number of interactive performer flows and the amount of application processing required by the MM service graph, whereas increasing $A$ increases the number of downstream delivery flows. 

\begin{figure*}[htbp]
    \centering

    \includegraphics[width=0.88\textwidth]{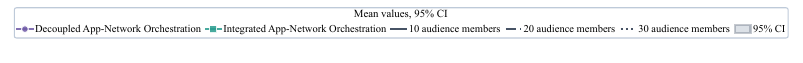}

    \begin{subfigure}[t]{0.32\textwidth}
        \centering
        \includegraphics[width=\linewidth]{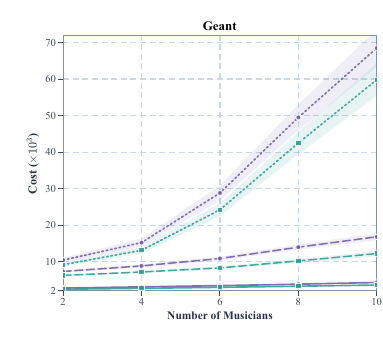}
        \caption{Overall resource cost (GEANT)}
        \label{fig:cost-geant}
    \end{subfigure}
    \hfill
    \begin{subfigure}[t]{0.32\textwidth}
        \centering
        \includegraphics[width=\linewidth]{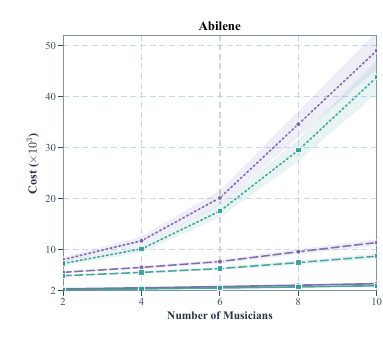}
        \caption{Overall resource cost (Abilene)}
        \label{fig:cost-abilene}
    \end{subfigure}
    \hfill
    \begin{subfigure}[t]{0.32\textwidth}
        \centering
        \includegraphics[width=\linewidth]{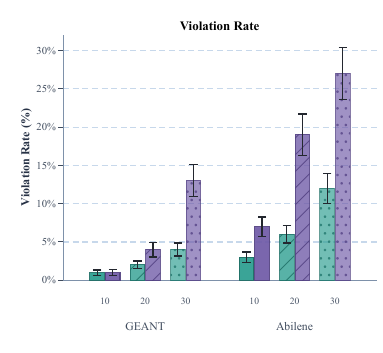}
        \caption{QoS Violation Rate.}
        \label{fig:violation-rate}
    \end{subfigure}

    \caption{Comparison of orchestration strategies in terms of cost and violation rate. 
    }
    \label{fig:cost-violation-comparison}
\end{figure*}

\subsection{Results and Discussion}
\label{subsec:results-discussion}

\Cref{fig:cost-geant}--\Cref{fig:cost-abilene} compare decoupled orchestration and integrated app-network orchestration in terms of overall cost of resources under varying number of musicians in three cases considering 10, 20 and 30 audience members. Higher number of musicians and audience members reflect a larger session size.
As expected, in all cases, and across both networks, greater session size incurs higher costs. This is due to tighter latency constraints and higher traffic volumes. More importantly, in all cases, integrated app-network orchestration attains lower costs with respect to decoupled orchestration. For instance, on GEANT (\Cref{fig:cost-geant}), the integrated orchestration attains lower cost than decoupled orchestrations ranging from 12\% to 27\%, depending on the audience size. Gains are more pronounced for larger session sizes (e.g., 8 or 10 musicians and 20 or 30 audience members). Similar trends are observed on Abilene (\Cref{fig:cost-abilene}), and, across all evaluated configurations and both topologies, integrated orchestration achieves an average cost reduction of approximately 14.9\%, with a maximum observed reduction of 27.4\% (27.4\% on GEANT and 23.3\% on Abilene), confirming that the benefits are not topology-specific.

It is interesting to note that the cost reductions that integrated app-network orchestration attains with respect to decoupled orchestration comes at a significant reduction in violation rates (as shown in \Cref{fig:violation-rate}, which reports the latency violation rate of both orchestration modes across both network topologies, with $M=4$). The results show that app-integrated orchestration reduces QoS violation rate by up to 70\%. For instance, for moderate audience sizes, violations are reduced by up to 50--70\%, and improvements remain substantial for larger session size. As previously mentioned, these gains are achieved alongside cost reductions, indicating that application-integrated slicing improves both efficiency and reliability rather than trading one for the other.

The observed gains stem from the ability of application-integrated slicing to coordinate placement, routing, and resource allocation across user roles and application functions. In contrast, decoupled orchestration treats application placement and network allocation as separate decisions, which limits its ability to capture the tight coupling between performer-side low-latency flows, audience-side bandwidth-intensive delivery, and cross-modal synchronization constraints.

Concluding, the results demonstrate that application-integrated slicing provides two key benefits: lower resource cost and improved service feasibility. While differences are modest under light load, they become substantial as the number of performers and audience members grows. 
Integrated orchestration enables coordinated placement, routing, and resource allocation decisions across application and network domains, resulting in lower resource consumption and improved QoS compliance. \emph{This suggests that future 6G orchestration frameworks can attain such benefits treating application and network resources as a unified optimization space.}

\section{From Service-Centric to Application-Integrated Slicing}

The results suggest that emerging immersive services challenge the traditional service-centric slicing paradigm. While network slicing has proven effective for resource isolation and KPI enforcement, future applications may involve heterogeneous user roles within the same service instance. A single slice-level KPI profile is therefore insufficient to capture the diverse latency, bandwidth, reliability, and synchronization requirements of all users. Future 6G architectures should complement service-centric slicing with application-aware orchestration mechanisms that jointly manage communication and computing resources across multiple user classes.

Current 3GPP slicing frameworks define slices as logical networks tailored to service requirements, with orchestration mechanisms allocating resources to satisfy slice-level KPIs~\cite{3gpp23501}. Although QoS flows and policy-based management support traffic differentiation, orchestration remains largely driven by service-level objectives and network-centric abstractions.

The MM highlights the limitations of this model. Within the same session, performers require ultra-low latency and tight synchronization, whereas audience members primarily require bandwidth-intensive immersive delivery. At the same time, both groups share application functions, network functions, and communication resources. Treating application placement and network allocation independently therefore fails to capture the coupling between user roles, shared functions, routing paths, and end-to-end constraints.

Application-integrated slicing addresses this limitation by modeling application microservices and network functions within a unified end-to-end service graph. Rather than associating a single KPI set with an entire service, orchestration can explicitly account for differentiated user roles, heterogeneous flows, and shared processing functions. Architecturally, this should be viewed as an evolution of network slicing rather than a replacement: existing slicing mechanisms continue to provide isolation and service assurance, while application-integrated slicing adds the semantic awareness required to optimize complex multi-user services across communication and computing domains.

Several challenges remain. First, \emph{scalability} becomes critical because joint application--network optimization increases orchestration complexity, motivating decomposition, hierarchical control, and AI-assisted approaches. Second, \emph{cross-domain interfaces} are needed to expose application requirements, user roles, and service objectives to network management systems while preserving interoperability. Third, \emph{security and trust} must be addressed, since richer application semantics improve optimization opportunities but also increase privacy, confidentiality, and attack-surface risks.

\section{Conclusion}
Emerging immersive services such as the Musical Metaverse challenge service-centric slicing and decoupled orchestration, since heterogeneous user roles may impose distinct latency, bandwidth, and synchronization requirements within the same service instance. To address this limitation, we introduced \emph{application-integrated slicing}, where application functions, network functions, routing, and communication/computation resources are jointly orchestrated in a unified service model. Numerical results show that this approach reduces both resource cost and KPI violations, with larger gains as the service size increases. These findings suggest that future 6G orchestration should move beyond purely network-centric abstractions and treat application and network resources as a unified optimization space for complex, interactive, and heterogeneous services.



\bibliographystyle{IEEEtran}
\bibliography{refs}

\end{document}